\documentclass[a4paper,11pt]{article}
\pdfoutput=1
 
\usepackage{jinstpub}
\usepackage{subcaption}
\usepackage{makecell}
\notoc
 
\title{\boldmath Spatial resolution for gamma ray interactions in stacks of monolithic silicon sensors}
 
\author[a,1]{L.~Terenzi,\note{Corresponding author.}}
\author[b]{S.~Garbolino,}
\author[a]{E.~Rieger,}
\author[a,c]{M.~Persson,}
\author[a]{M.~Yveborg~Tamm,}
\author[b]{M.~Da~Rocha~Rolo,}
\author[d]{L.~Pancheri,}
\author[a,c]{and M.~Danielsson}

\affiliation[a]{Department of Physics, KTH Royal Institute of Technology, SE-106 91 Stockholm, Sweden}
\affiliation[b]{INFN Sezione di Torino, Italy}
\affiliation[c]{MedTech-Labs, BioClinicum, Karolinska University Hospital, SE-171 64 Solna, Sweden}
\affiliation[d]{Department of Industrial Engineering, University of Trento, Via Sommarive 9, 38123 Trento, Italy}
 
\emailAdd{lucat@kth.se}
 
\abstract{

We are developing a new implementation of Compton imaging for nuclear medicine using a large volume of stacked monolithic silicon sensors. In this work we investigate how the spatial resolution is impacted by the pixel size, an important input to the design of the sensors. In general CMOS design is less challenging with larger pixels leaving more space for analog and digital electronics. On the other hand, spatial resolution is one of the key parameters that will impact the performance of the Compton imaging system. It is also important to consider the range of the recoil electron produced in Compton or photoelectric interactions, which impacts how accurately the interaction point can be estimated. The achievable resolution was evaluated with two interaction position reconstruction algorithms: one based on a Gaussian fit of the detected charge and a second based on the characteristic energy deposition along the electron track, with the former performing better at lower energies. Monte Carlo simulations were performed for 140\,keV and 511\,keV sources embedded in a water phantom, with pixel pitches ranging from 25\,$\mu$m to 250\,$\mu$m.  Energy-averaged in-plane resolution degraded from 8\,$\mu$m to 94\,$\mu$m as pixel size increased. The out-of-plane resolution was 216\,$\mu$m at low energies, limited by sensor thickness, and improved to 155\,$\mu$m at higher energies. The results suggest that pixel sizes in the order of 100\,$\mu$m can achieve spatial resolutions on the order of tens of micrometers.

}
 

\begin{document}
\maketitle
\flushbottom
 
% ---------------------------------------------------------------
\section{Introduction}
\label{sec:intro}
% ---------------------------------------------------------------
 
Silicon detectors have become a cornerstone of modern radiation detection due to their excellent charge collection properties, technological maturity, and scalability. They are widely used for charged particle tracking in high-energy physics, X-ray and photon counting applications, and high-resolution imaging systems \cite{moser2009,greffier2025}. Our group has previously demonstrated high-resolution photon-counting performance in silicon-based devices~\cite{sundberg2021,danielsson2021}, highlighting the potential of silicon technology for precise interaction localization in medical imaging.
Recent advances in monolithic silicon technology further enhance these capabilities by integrating the sensing volume and readout electronics within a single substrate, yielding compact, low-power detector modules with  straightforward tiling~\cite{turchetta2001,darocharolo2025}.
These characteristics are particularly attractive for Compton imaging, which reconstructs the incident photon direction from the kinematics of multiple interactions inside the detector~\cite{caravaca2022,krimmer2015,yoon2020,andreyev2010}, eliminating the need for a mechanical collimator, and offering higher sensitivity and resolution over conventional SPECT systems based on pin-hole collimators~\cite{vanaudenhaege2015}. Achieving high Compton imaging performance critically depends on accurate localization of each interaction point and precise energy measurement, where high spatial resolution and low electronic noise are essential to minimize angular uncertainty~\cite{kim2024,vetter2007}.

We are developing a stacked monolithic silicon detector with full 3D position sensitivity for nuclear medicine applications designed to handle complex interaction chains, including events with more than two interactions or incomplete energy deposition, thereby increasing usable event statistics compared to conventional scatter and absorber two-detector Compton cameras. Moreover, no distinction between scatterer and absorber is present in the detector geometry, as both Compton scattering and photoelectric absorption can take place anywhere within the detector volume. Accurate localization of each interaction point is a critical requirement, as the spatial resolution of the reconstructed interaction positions directly influences the angular resolution and ultimately the image quality of the Compton camera. Charge sharing between neighboring pixels, arising from diffusion of the charge cloud, can be exploited to achieve sub-pixel localization precision~\cite{sundberg2021}, where smaller pixel pitches allow finer spatial sampling and thus improved resolution. However, reducing the pixel pitch comes with trade-offs: smaller pixels might lead to an increased complexity of the pixel electronics, since the silicon area available for the pixel circuitry would be reduced. Even assuming a proportional decrease of the front-end amplifier power budget with a smaller capacitance at the input node, the overall power consumption of the detector could increase along with number of readout channels.
The intrinsic interaction localization limit of the detector geometry will be investigated for 511\,keV and 140\,keV photon sources across multiple pixel size configurations providing design input for the readout electronics based on previous work for the ARCADIA CMOS sensors~\cite{darocharolo2025} and benchmarking the performance against state-of-the-art Compton imaging detectors, some of which are reported in table~\ref{tab:comparison}.

\begin{table}[htbp]
\centering
\caption{Position resolution for existing gamma-ray detectors used in Compton imaging. The reported resolutions refer to the scatterer.}
\label{tab:comparison}
\begin{tabular}{lcl}
\hline
\textbf{Reference} & \textbf{Position resolution ($\mu$m)} & \textbf{Application (material)} \\
\hline
Caffrey et al.\ 2021~\cite{caffrey2021}    & 5000                                    & Industrial applications (Si) \\
\makecell[lt]{Du et al.\ 2001~\cite{du2001}} & \makecell[ct]{1000 (in-plane) \\ 500 (out-of-plane)} & \makecell[lt]{General Compton imaging (CZT)} \\
Griffiths et al.\ 2026~\cite{griffiths2026}& 1000 (3D)                               & General Compton imaging (Ge) \\
Park et al.\ 2007~\cite{park2007}          & 700 (in-plane)                          & Nuclear medicine (Si-BGO) \\
Takeda et al.\ 2009~\cite{takeda2009}      & 400 (in-plane)                          & Astrophysics/Nuclear medicine (Si) \\
Yabu et al.\ 2022~\cite{yabu2022}          & 250 (in-plane)                          & Nuclear medicine (Si-CdTe) \\
Turecek et al.\ 2018~\cite{turecek2018}          & 55 (in-plane)                          & Nuclear medicine (Si-CdTe) \\
\hline
\end{tabular}
\end{table}
 
% ---------------------------------------------------------------
\section{Methods}
\label{sec:methods}
% ---------------------------------------------------------------
 
\subsection{Detector geometry}
\label{sec:geometry}
 
For this study, we investigated a detector with 67 layers of silicon sensors for a total volume of $20\,\text{cm} \times 20\,\text{cm} \times 5.025\,\text{cm}$ (no spacing is introduced between sensors). Four different pixel sizes were simulated, as described in table~\ref{tab:pixel_sizes}, where the choice of the electrodes number and sizes derives from preliminary studies to reduce collection time while maintaining a low capacitance.
 
\begin{table}[htbp]
\centering
\caption{Pixel configurations tested. All sensors have the same thickness of 750\,$\mu$m.}
\label{tab:pixel_sizes}
\begin{tabular}{cccc}
\hline
\textbf{Pixel size} ($\mu$m) & \textbf{Electrodes} & \textbf{Electrodes pitch} ($\mu$m) & \textbf{Electrode size} ($\mu$m) \\
\hline
$25\times25$ & 1 & $25\times25$ & $6\times6$ \\
$50\times50$ & 1 & $50\times50$ & $6\times6$ \\
$100\times100$ & 4 & $50\times50$ & $6\times6$ \\
$250\times250$ & 4 & $125\times125$ & $14\times14$ \\
\hline
\end{tabular}
\end{table}
 
\begin{figure}[htbp]
\centering
\includegraphics[width=0.75\textwidth]{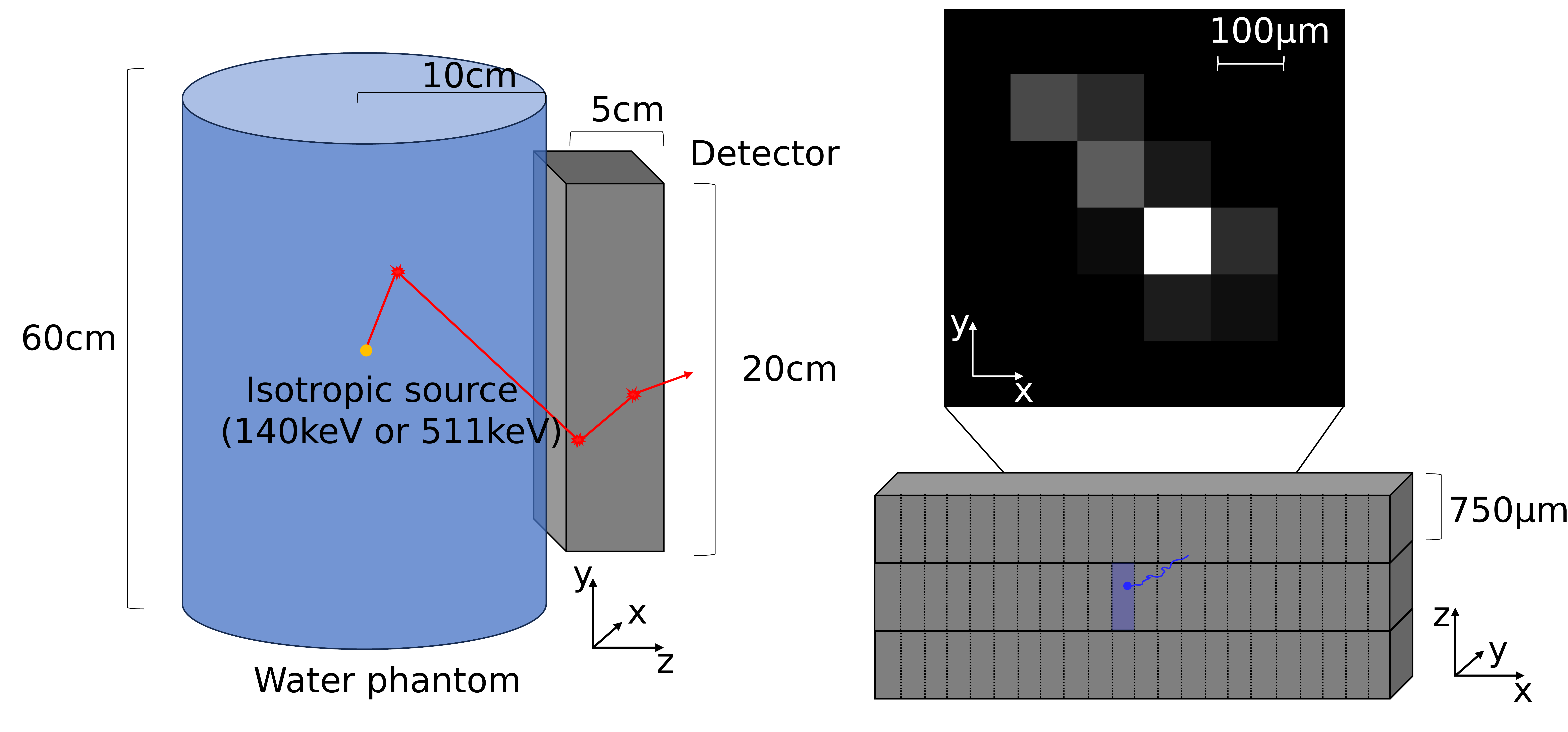}
\caption{An illustration of the simulated set-up (left) and an example of an electron track (right) for the case of $100\mu$m pixel size.}
\label{fig:setup}
\end{figure}

This work defines wafer depth as the $z$ direction, while the squared pixels are aligned with the $x$-$y$ plane. Point-like isotropic photon sources of energy 140\,keV or 511\,keV were used, located at position $(0,0,0)$\,cm, while the detector surface was placed 10\,cm away from it along the $z$ direction. To mimic a biologically relevant environment, the source was embedded in a cylindrical volume of water, 10\,cm in radius and 60\,cm in height, with its longitudinal axis aligned parallel to the plane of the detector, as depicted in figure~\ref{fig:setup}.
 
\subsection{Photon source and electron tracks}
\label{sec:source}
 
Photon interactions and electron tracks were simulated with the Allpix$^2$ framework, which uses the GEANT4 toolkit~\cite{spannagel2018}, with subsequent data processing carried out in Python. Allpix$^2$ records the energy deposition and positions of the photon interactions in the detector, simulating also secondary particles generated such as the recoil electrons, Auger processes, ionization electrons, fluorescence and Bremsstrahlung photons. For the electron tracks, the software can also record the energy deposited as they travel, creating electron-hole pairs and propagating them using a weighting and electric field provided by the user. Therefore, we can simulate the full chain from the photon emission to the charge collected by the pixels, which is then used in post-processing to estimate the interaction positions.
 
A total of 1\,000\,000 photon emissions were simulated for each pixel geometry and source energy, leading to around 280\,000 interactions for each set of parameters.
 
\begin{figure}[htbp]
\centering
\includegraphics[width=0.75\textwidth]{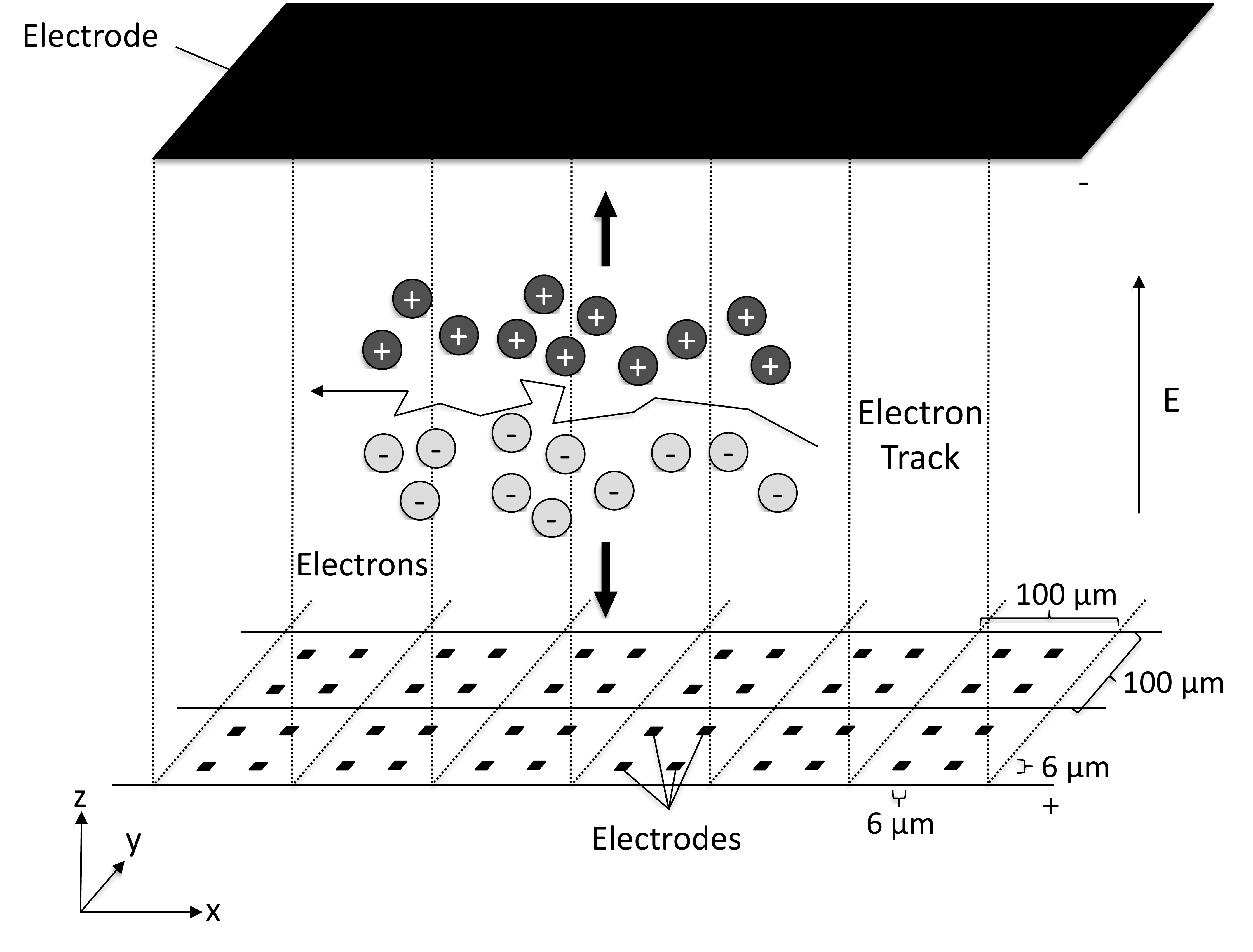}
\caption{Visualisation of the sensor geometry used for simulations for the case of $100\mu$m pixels.}
\label{fig:geometry}
\end{figure}
 
\subsection{Signal induction}
\label{sec:signal}
 
The electron-hole pairs generated by the electron track are separated by a bias voltage of $-200$\,V in the $z$-direction along the wafer thickness. The bias voltage makes the electrons and holes drift in opposite directions towards the electrodes (see figure~\ref{fig:geometry}). Diffusion occurs during the drift, spreading the initial charge cloud in the $x$- and $y$-direction. The drift-diffusion model implemented in Allpix$^2$ accounts for thermal diffusion and drift under the applied electric field, but does not include Coulomb repulsion between charge carriers during propagation~\cite{spannagel2018}. The induced current on electrode $k$ follows the Shockley--Ramo theorem:

\begin{equation}
  i_{k}(t) = e\,\vec{E}_{w,k}\!\left(x(t),y(t),z(t)\right)\vec{v}(t) \,,
  \label{eq:ramo}
\end{equation}
where $e$ is the elementary charge, $\vec{v}(t)$ is the carrier velocity, and $\vec{E}_{w,k}$ is the weighting field for electrode $k$.The weighting field is computed in COMSOL Multiphysics from the weighting potential $W$, obtained by solving the Laplace equation:
\begin{equation}
  \nabla^{2} W = 0 \,,
  \label{eq:laplace_W}
\end{equation}
with boundary conditions $W=1$ on the target electrode $k$ and $W=0$ on all others. The weighting field is then obtained as:
\begin{equation}
  \vec{E}_{w,k}\!\left(x(t),y(t),z(t)\right) = -\nabla W \,,
  \label{eq:laplace}
\end{equation}
This theoretical framework is based on previous works~\cite{sundberg2021} with an extension to 3 dimensions.
 
\subsection{Readout electronics and noise model}
\label{sec:electronics}
 
The analog front-end response was modeled by integrating the induced current to obtain the collected charge per pixel. Pixel signal fluctuations were modeled with a Normal distribution with a standard deviation of 90 electrons, comparable to previous results for a similar ASIC~\cite{olave2018}. An energy threshold of approximately 2 keV was then applied to the pixel data. The Gaussian noise contribution from the readout electronics, together with the pixel threshold, were not applied during the Allpix$^2$ simulations, but instead in post-processing to allow easier control over the parameters without the need to rerun the simulations. No discretizations or gain functions were applied to assess the fundamental resolution of the proposed geometries, and dark counts (pixels activated by noise) were also not simulated for simplicity.
 
\subsection{Data processing and interaction position reconstruction}
\label{sec:reconstruction}
 
The raw data of the true photon interaction positions and deposited energies are saved, together with the relative pixel activations and collected charges. The noise is then added to the charge values collected by the pixels; the charge is converted to energy assuming a 3.6\,eV silicon ionization energy per electron-hole pair, and the pixel threshold is applied. For each event, the remaining activated pixels that are spatially adjacent are grouped into clusters using a $3 \times 3 \times 3$ connectivity mask (27-voxel neighborhood). This procedure groups neighboring activated pixels into a single cluster, which is assumed to originate from the same interaction. If pixels activations deriving from two separate interactions are mistakenly grouped together due to their spatial proximity, the interactions are discarded. The resulting clusters are then projected into the $x$-$y$ plane if they extend to an adjacent sensor, to apply the in-plane position reconstruction algorithms. For low energy, the charge cloud can be approximated by a sphere, while for higher energies the electron tracks become more extended and therefore spread over multiple pixels. We accordingly applied two different algorithms adapted to either case. The first algorithm is referred to as \emph{Gaussian fit}, while the second as \emph{Gridsum~$N$}. Both algorithms are applied over the full energy range of interactions to study their relative and absolute in-plane resolution performance.
 
\paragraph{Algorithm~1, Gaussian fit.}
The pixel values in keV obtained from the data processing are used to estimate the location of the interaction. This is accomplished by summing the values over one dimension for the $x$- and $y$-directions separately, such that two one-dimensional projections are obtained. For both dimensions, a Gaussian normal function
\begin{equation}
  f(x) = a\exp\!\left(-\frac{(x-b)^{2}}{2c^{2}}\right)
\end{equation}
is fitted to the projection. Due to charge sharing between pixels, the mean $b$ of the fitted function corresponds to the location of the interaction. If a one-dimensional projection presents 4 or fewer active pixels, a weighted average is used instead:
\begin{equation}
  \bar{x} = \displaystyle\sum_{i=1}^{n} w_{i}\,x_{i} \,,
  \qquad
  w_{i} = \frac{E_{i}}{\displaystyle\sum_{i=1}^{n} E_{i}} \,,
  \label{eq:wavg}
\end{equation}
where $x_{i}$ are the pixel centre coordinates and $E_{i}$ is the signal in keV collected by pixel $i$. This approach avoids cases in which the Gaussian fit would be unreliable or fail.
 
\paragraph{Algorithm~2, Gridsum~$N$.}
Higher-energy electrons produce longer tracks, with energy deposition patterns exhibiting a more defined Bragg peak: as the electron decelerates, the rate of energy loss increases. Therefore, the least energy is deposited near the beginning of the track, where the electron velocity is highest~\cite{plimley2011, plimley2013}. The Gridsum~$N$ algorithm exploits this by identifying the pixel center with the lowest total charge collected in its surrounding $N\times N$ neighborhood, which corresponds to the beginning of the track and thus the photon interaction position.

\begin{figure}[htbp]
\centering
\includegraphics[width=0.55\textwidth]{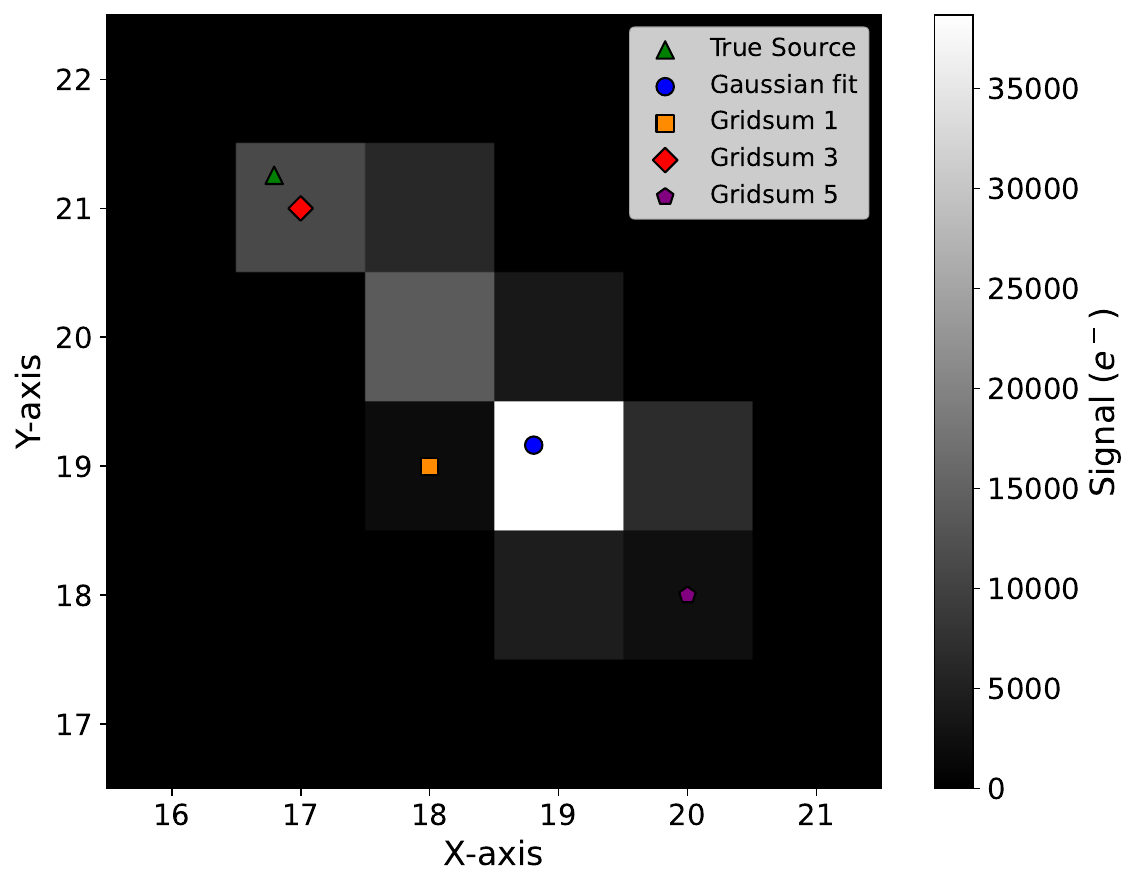}
\caption{An example of the algorithms' application for 100\,$\mu$m pixel pitch. The Gaussian fit is biased by the high charge accumulation ($>$35\,000\,e$^{-}$) around the center of the track. Gridsum~1 selects the pixel collecting the least charge, while Gridsum~5, having too large a grid, does not properly capture the Bragg-peak distribution.}
\label{fig:algorithm}
\end{figure}

For each interaction, let $S(x_i, y_i)$ denote the discrete energy map of the activated pixels projected onto the $x$-$y$ plane, where each entry corresponds to the charge collected by the pixel whose center is at $(x_i, y_i)$. Let $N$ be the side length of the square summation window (with $N$ an odd integer); the Gridsum~$N$ matrix is defined as
\begin{equation}
  G_{N}(x_i,y_i) = \sum_{|j|\leq r}\sum_{|k|\leq r} S(x_i+j,\,y_i+k) \,,
  \qquad
  r = \frac{N-1}{2} \,,
  \label{eq:gridsum}
\end{equation}
where $j$ and $k$ are integers indexing the neighboring pixels within the window. Let $P = \{(x_i, y_i)\}$ denote the set of activated pixel centers in the $x$-$y$ projection. The interaction position is then estimated as the pixel center minimizing $G_{N}$:
\begin{equation}
  \left(\hat{x_i},\hat{y_i}\right) = \underset{(x_i,y_i)\in P}{\arg\min}\; G_{N}(x_i,y_i) \,,
  \label{eq:argmin}
\end{equation}
with ties resolved by selecting the pixel with minimal $S(x_i,y_i)$. An example of the algorithm's application is shown in figure~\ref{fig:algorithm}.

To compute the in-plane spatial resolution of the algorithms, the radial distance between the estimated interaction point and the true one was computed as $\text{error} = \|\vec{r}_{\,\text{true}} - \vec{r}_{\,\text{estimated}}\|$, where $\vec{r} = (x, y, z)$. The results were binned on the interaction deposited energies, for a total of 25 bins between 0\,keV and 550\,keV, and in each bin the root mean square (RMS) error was calculated.
Due to the low statistics for bins above the Compton edge, only results from bins whose center is below 341\,keV will be shown.

To estimate the $z$ interaction position, the pixel values in each cluster are summed over the $x$-$y$ plane to obtain a value per sensor. If the electron track activates only one sensor, the interaction position is estimated at half the sensor height. If two adjacent sensors are activated, the $z$ coordinate is estimated as the midpoint between the two layers of silicon (at their interface). This second case assumes that when an interaction results in the activation of two sensors, it must have taken place close to the interface between the two. In the studies, no interaction ever activated more than two adjacent sensors, and given the energy range of the electron tracks and the sensor thickness, this is highly unlikely to occur. The resolution is then estimated as the standard deviation (std) of the $\text{error} = {z}_{\,\text{true}} - {z}_{\,\text{estimated}}$.
 
% ---------------------------------------------------------------
\section{Results and discussion}
\label{sec:results}
% ---------------------------------------------------------------
 
\begin{figure}[htbp]
    \centering
    \begin{subfigure}{0.48\textwidth}
        \centering
        \includegraphics[width=\textwidth]{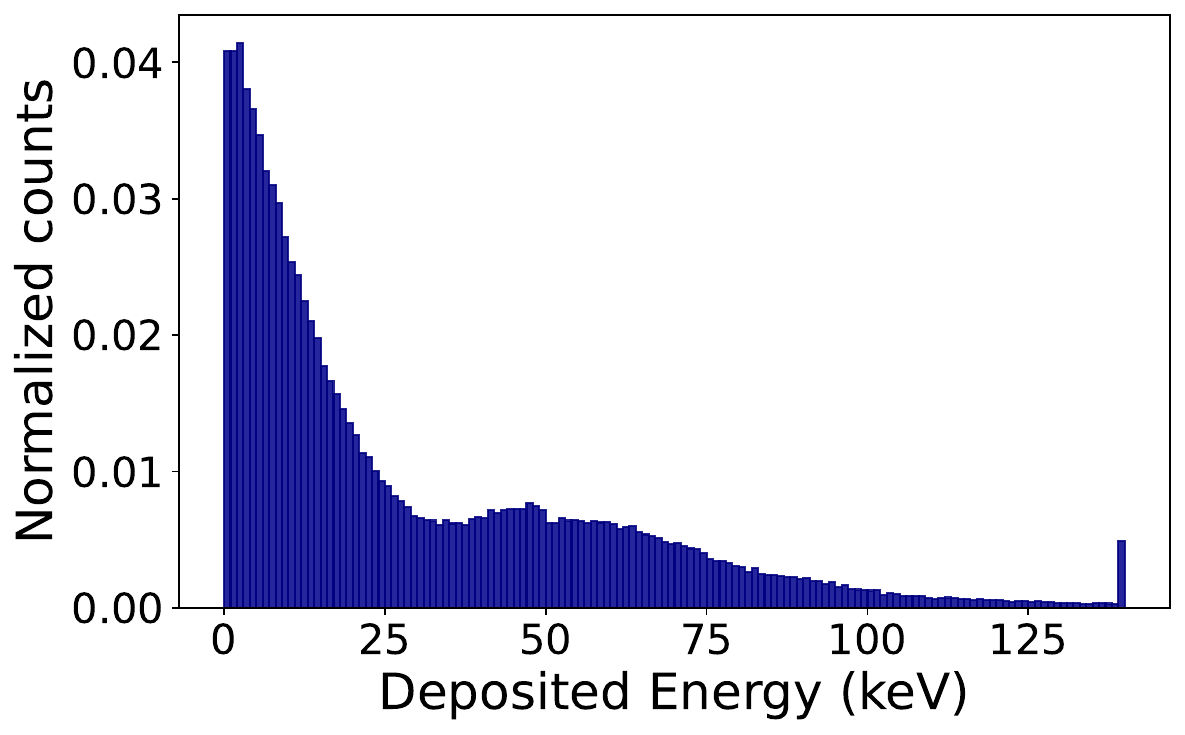}
        \caption{}
        \label{fig:spectrum_140}
    \end{subfigure}
    \hfill
    \begin{subfigure}{0.48\textwidth}
        \centering
        \includegraphics[width=\textwidth]{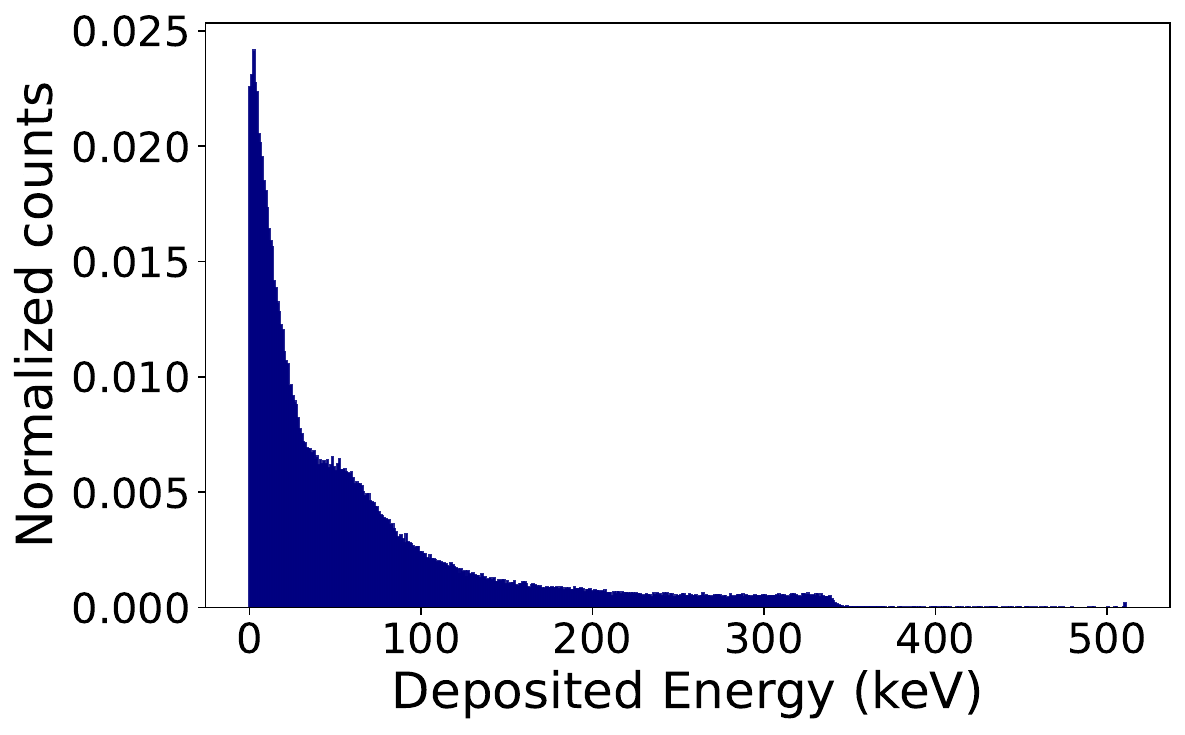}
        \caption{}
        \label{fig:spectrum_511}
    \end{subfigure}
    \caption{Energy spectra of deposited energies from photon interactions with the detector for a monochromatic source of 140\,keV (a) and 511\,keV (b). The bin width used in both cases is 1\,keV.}
    \label{fig:spectra}
\end{figure}

Figure~\ref{fig:spectra} shows the distributions of deposited energy in the detector interactions from photons simulated with both 140\,keV and 511\,keV source energies. In both cases, most of the interactions deposit low energy, roughly below 30\,keV, and only in rare cases does the deposited energy exceed the 511\,keV Compton edge at approximately 341\,keV. The following plots are derived from the interaction dataset generated using the 511\,keV source simulation. This approach is justified by the fact that the deposited energy spectrum corresponding to 140\,keV interactions lies entirely within the energy range covered by the 511\,keV simulation.
 
\begin{figure}[htbp]
    \centering
    \begin{subfigure}{0.48\textwidth}
        \centering
        \includegraphics[width=\textwidth]{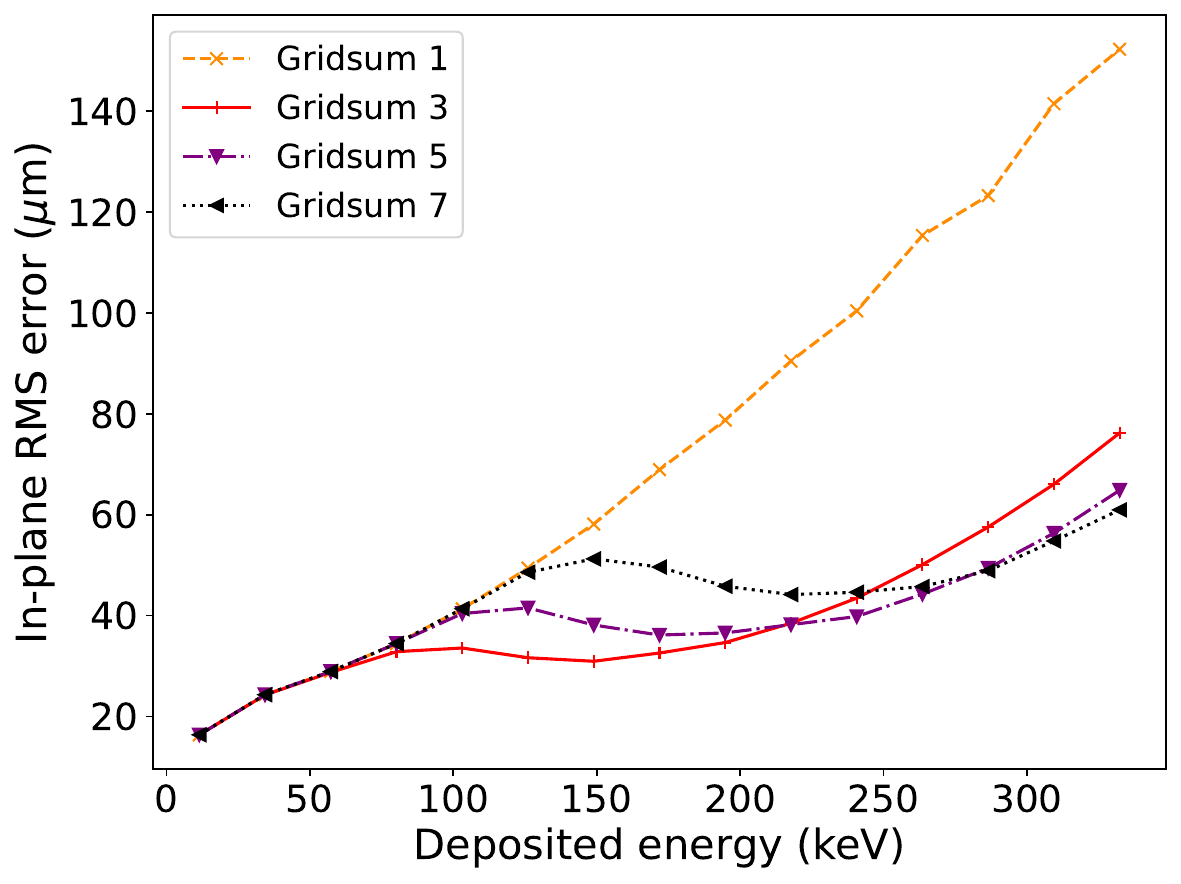}
        \caption{}
        \label{fig:gridsum_n_25um}
    \end{subfigure}
    \hfill
    \begin{subfigure}{0.48\textwidth}
        \centering
        \includegraphics[width=\textwidth]{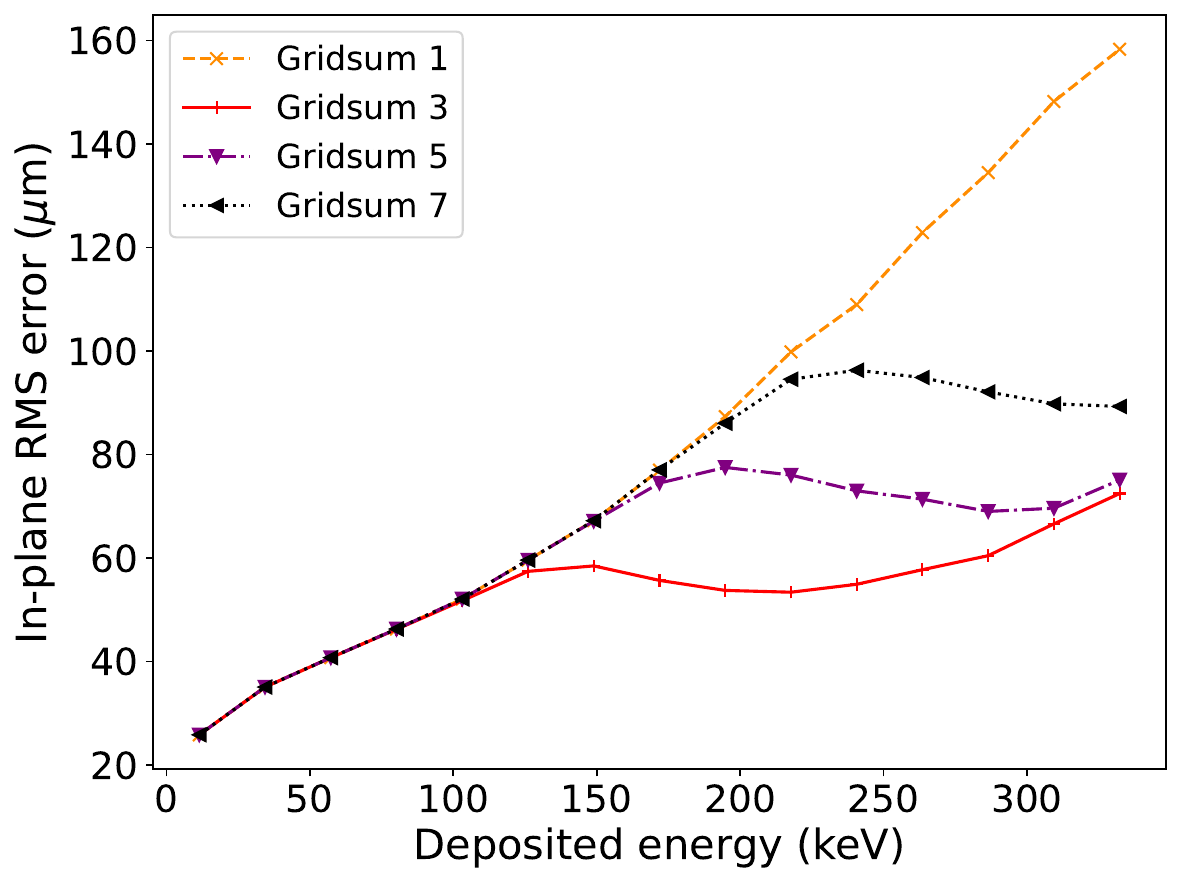}
        \caption{}
        \label{fig:gridsum_n_50um}
    \end{subfigure}
    \par\medskip
    \begin{subfigure}{0.48\textwidth}
        \centering
        \includegraphics[width=\textwidth]{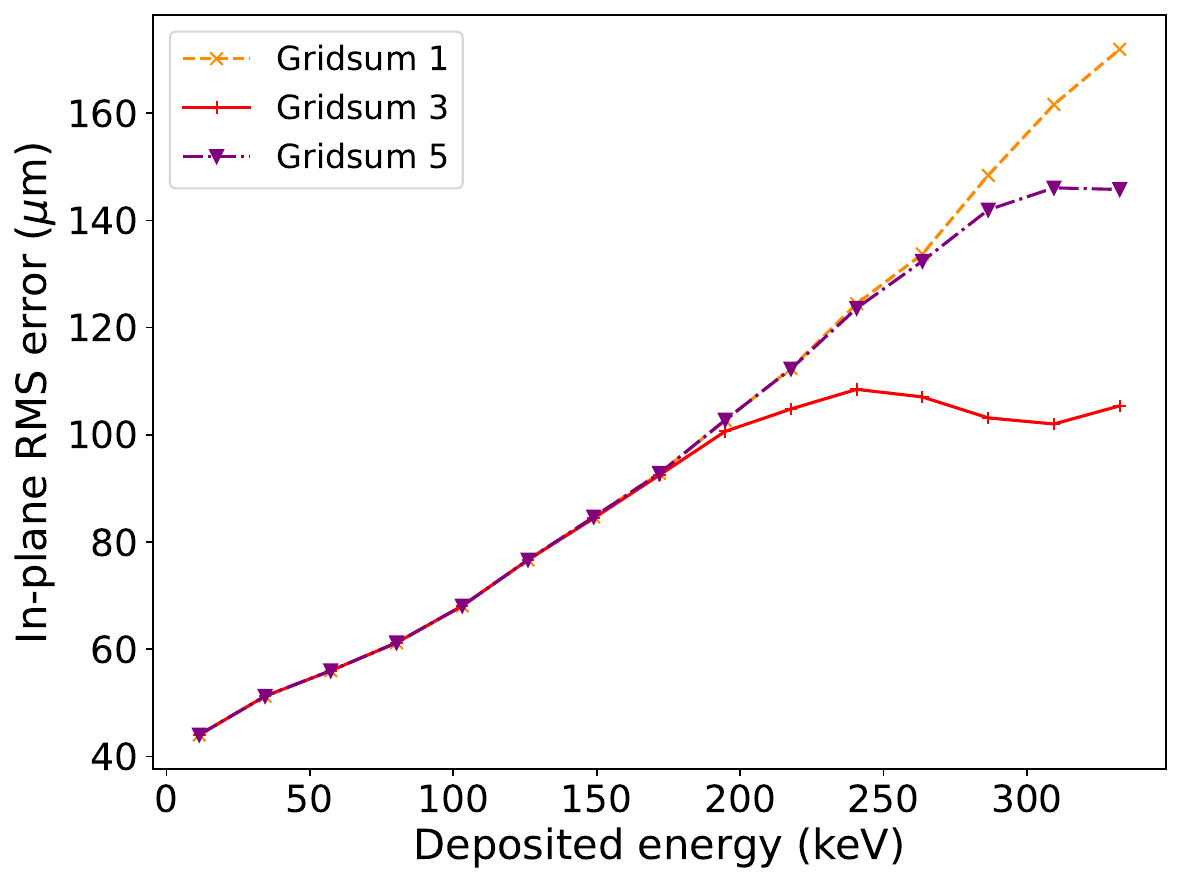}
        \caption{}
        \label{fig:gridsum_n_100um}
    \end{subfigure}
    \hfill
    \begin{subfigure}{0.48\textwidth}
        \centering
        \includegraphics[width=\textwidth]{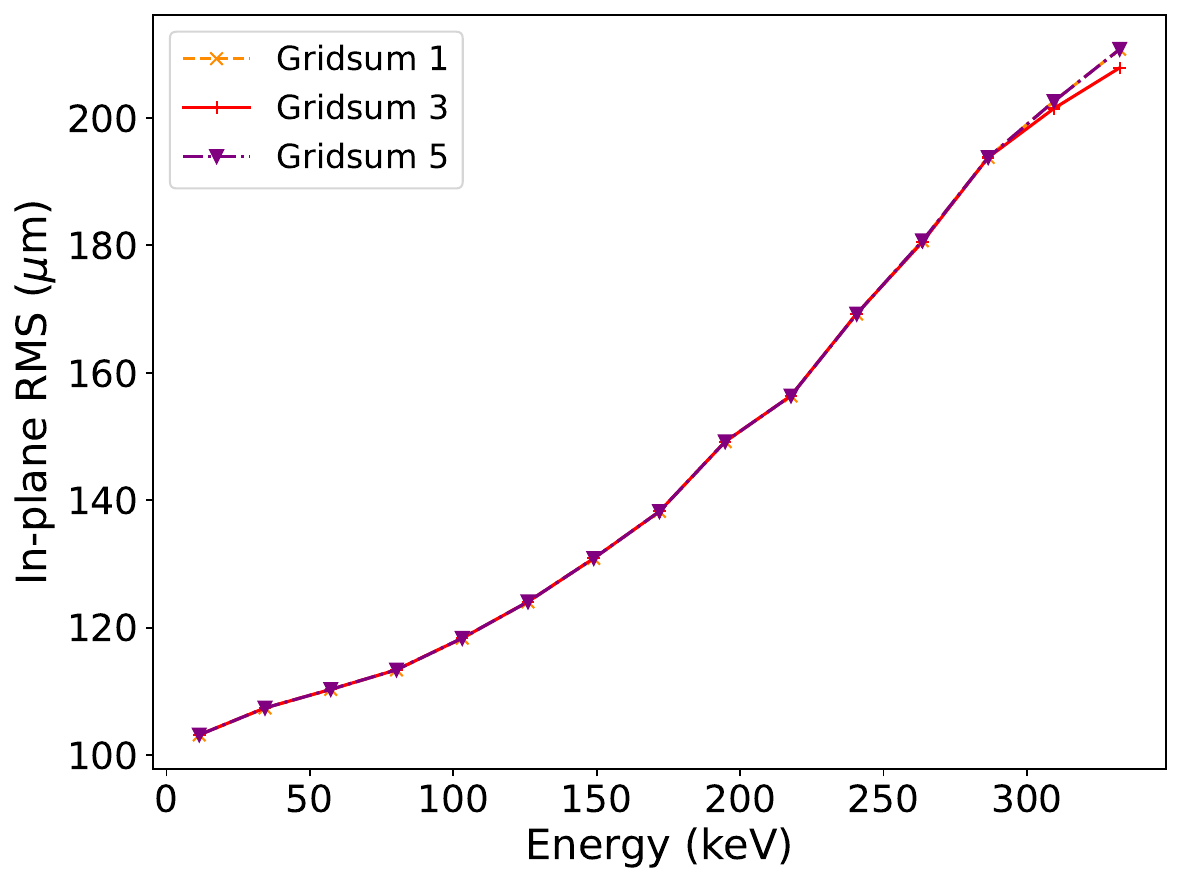}
        \caption{}
        \label{fig:gridsum_n_250um}
    \end{subfigure}
    \caption{In-plane RMS comparison between different $N$ values for $25\,\mu$m (a), $50\,\mu$m (b), $100\,\mu$m (c) and $250\,\mu$m (d) pixel pitches.}
    \label{fig:gridsum_n}
\end{figure}

Before comparing the algorithms, the grid size $N$ of the Gridsum algorithm was investigated to optimize the in-plane spatial resolution. Values between 1 and 7 were tested, where $N=1$ corresponds to simply choosing as the interaction position the center of the pixel with the lowest collected charge. For all pixel sizes investigated but the 25$\mu$m$\times$25$\mu$m one, Gridsum~3 outperformed all the other $N$ values across the spectrum of deposited energies until the Compton edge at 341\,keV as shown in figure~\ref{fig:gridsum_n} . For the smallest pixel size tested, due to the high number of activated pixels, there were different optimal $N$ values for the Gridsum algorithm at different energies as shown in figure~\ref{fig:gridsum_n_25um}.
 
Nonetheless, some similarities can be drawn from all cases: at low interaction energies, where only few pixels are activated, all Gridsum~$N$ algorithms converge to the same result, as the summed charge within any grid size is identical and the interaction position is assigned according to the draw rule described in section~\ref{sec:reconstruction}, leading to the same estimate. As the interaction energy increases and more pixels are activated, smaller grid sizes ($N$) become effective at isolating the Bragg peak region. However, beyond a certain energy threshold (figure~\ref{fig:gridsum_n_25um}), Gridsum~3 becomes susceptible to bias introduced by low-energy deposition regions along the middle of the track for example, which are spatially distant from both the Bragg peak and the true interaction position. Similar trends are visible for the other Gridsum~$N$ algorithms tested in figure~\ref{fig:gridsum_n_25um}. In figure~\ref{fig:gridsum_n_50um} and figure~\ref{fig:gridsum_n_100um} Gridsum~3 also exhibits local minima, but at higher energies due to the different pixel size. In general all Gridsum~$N$ algorithms tested showed a similar behavior at different energies depending on the pixel pitch and $N$ value. Ultimately, Gridsum~3 was identified as best performing for pixel pitches of $50\,\mu$m, $100\,\mu$m and $250\,\mu$m, while for the $25\,\mu$m configuration the best Gridsum~$N$ was determined independently for each energy bin (this case will be later referred to as Gridsum~$N*$).

\begin{figure}[htbp]
    \centering
    \begin{subfigure}{0.48\textwidth}
        \centering
        \includegraphics[width=\textwidth]{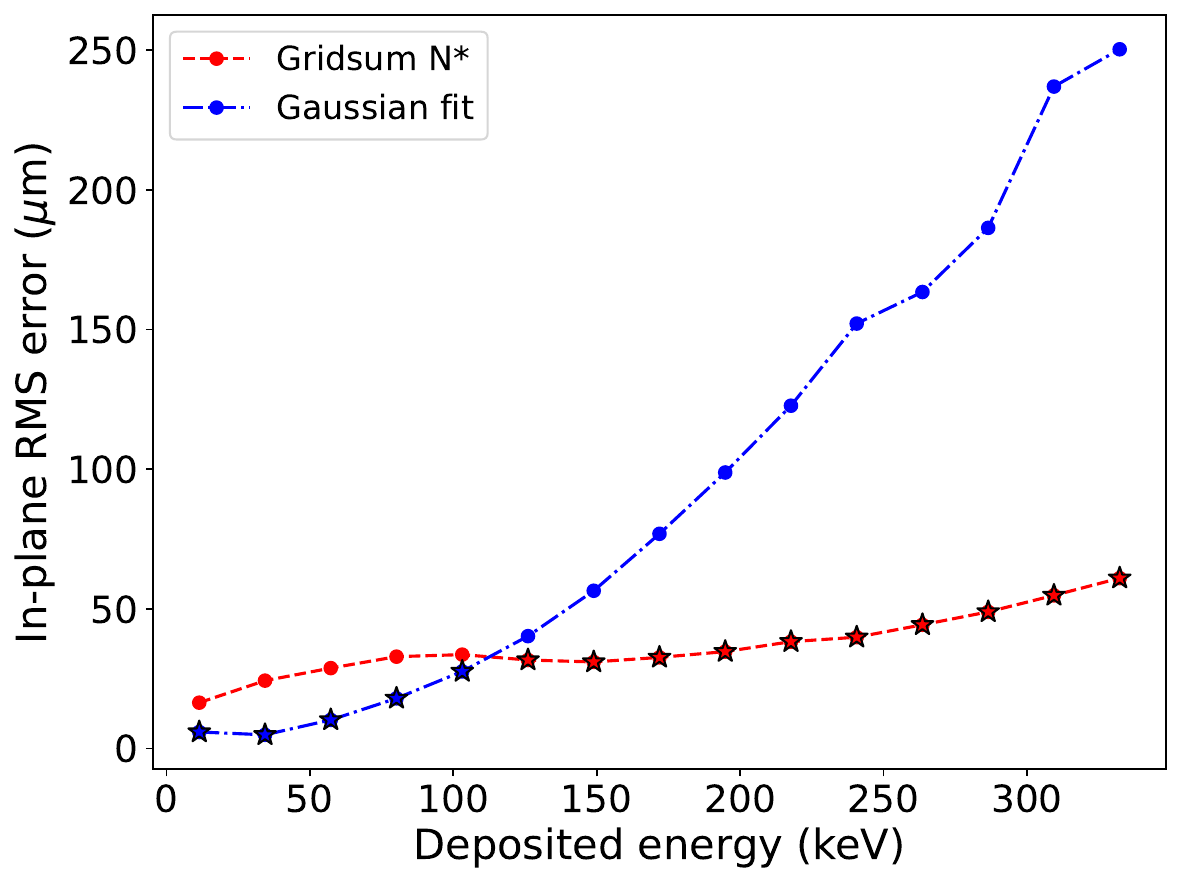}
        \caption{}
        \label{fig:grid_vs_gauss_25um}
    \end{subfigure}
    \hfill
    \begin{subfigure}{0.48\textwidth}
        \centering
        \includegraphics[width=\textwidth]{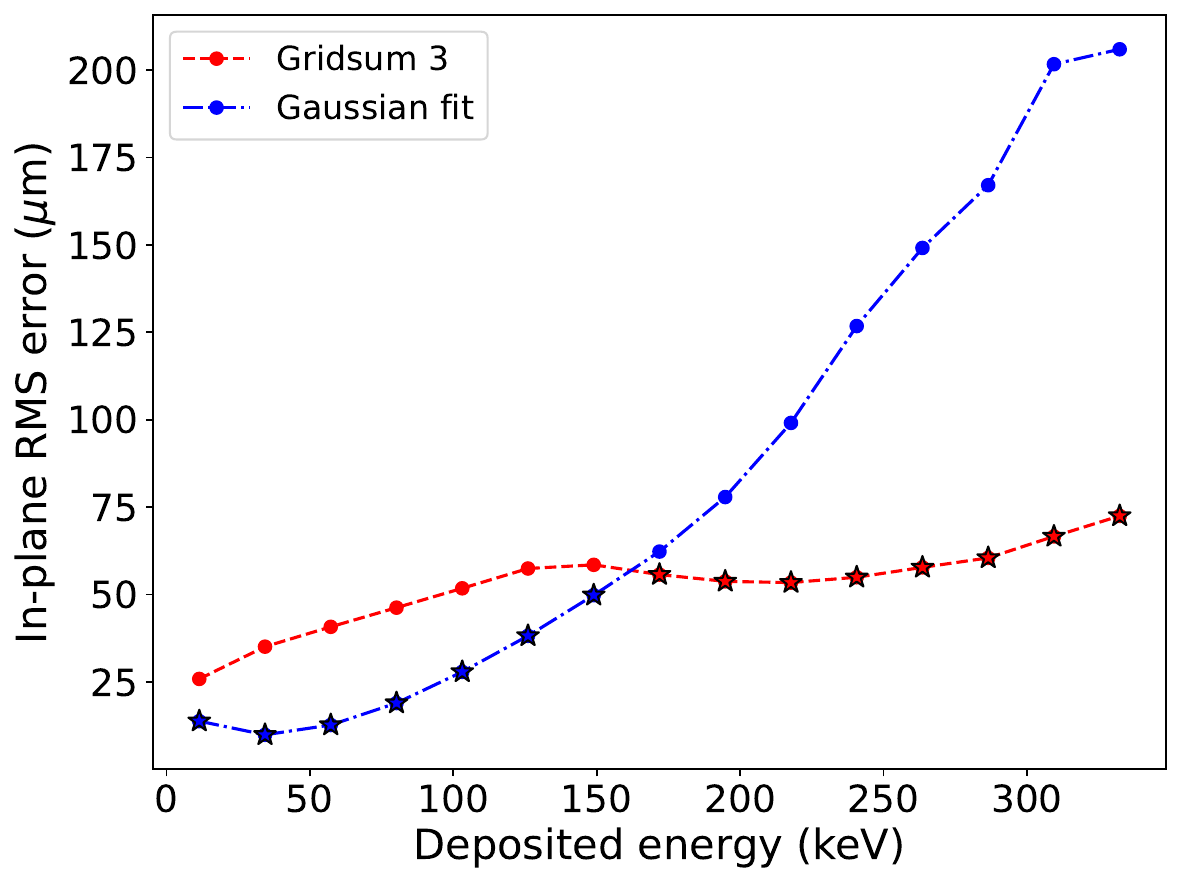}
        \caption{}
        \label{fig:grid_vs_gauss_50um}
    \end{subfigure}
    \par\medskip
    \begin{subfigure}{0.48\textwidth}
        \centering
        \includegraphics[width=\textwidth]{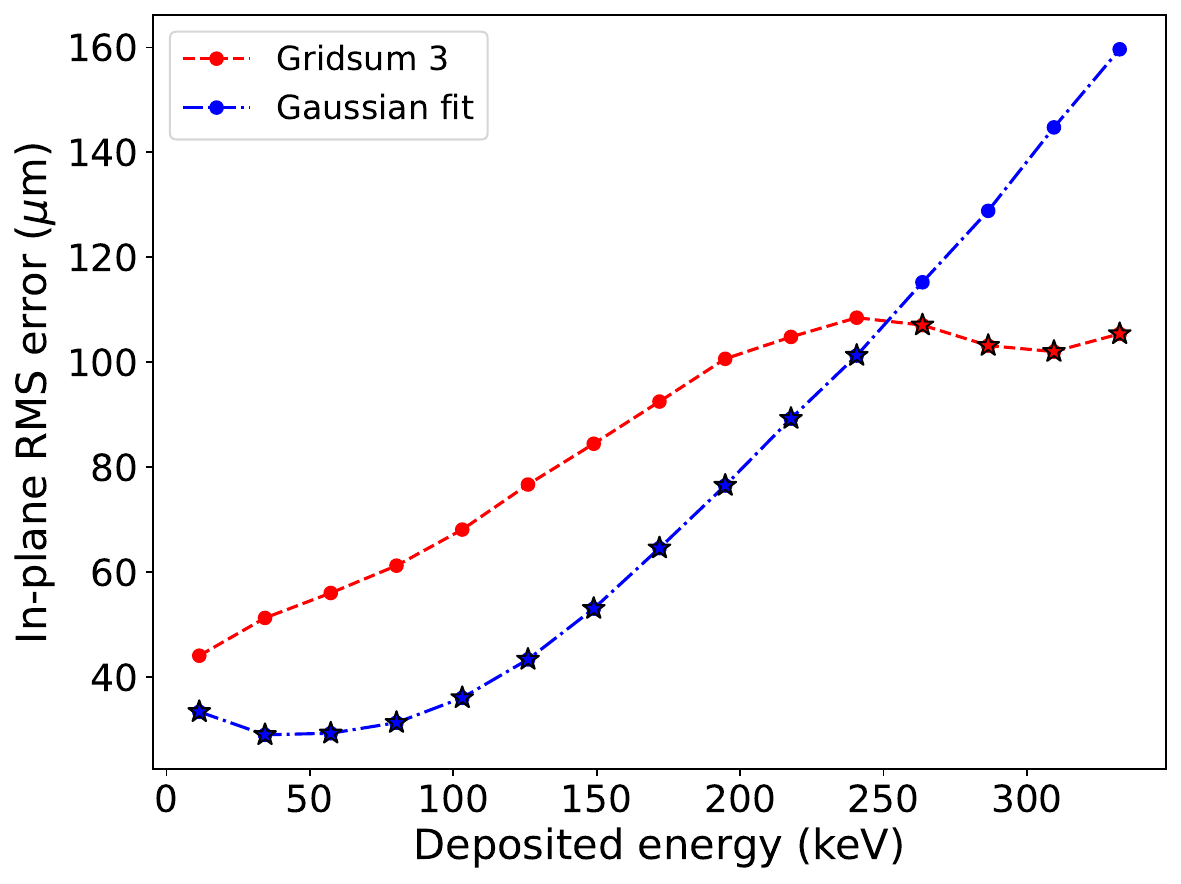}
        \caption{}
        \label{fig:grid_vs_gauss_100um}
    \end{subfigure}
    \hfill
    \begin{subfigure}{0.48\textwidth}
        \centering
        \includegraphics[width=\textwidth]{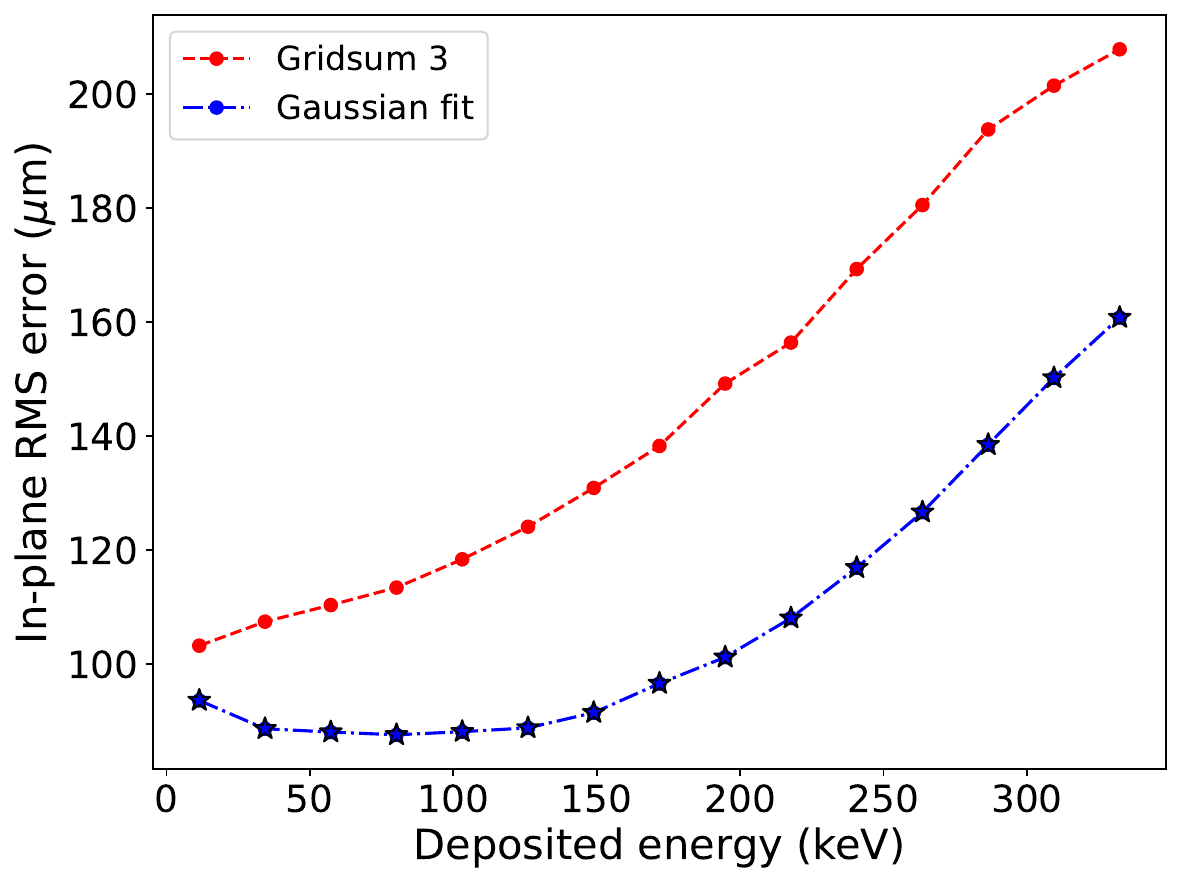}
        \caption{}
        \label{fig:grid_vs_gauss_250um}
    \end{subfigure}
    \caption{In-plane interaction position RMS error as a function of deposited energy for the Gridsum and Gaussian fit reconstruction methods for $25\,\mu$m (a), $50\,\mu$m (b), $100\,\mu$m (c) and $250\,\mu$m (d) pixel pitches. Star markers indicate the algorithm yielding the lowest position error in each energy bin.}
    \label{fig:grid_vs_gauss}
\end{figure}

\begin{figure}[htbp]
\centering
\includegraphics[width=0.55\textwidth]{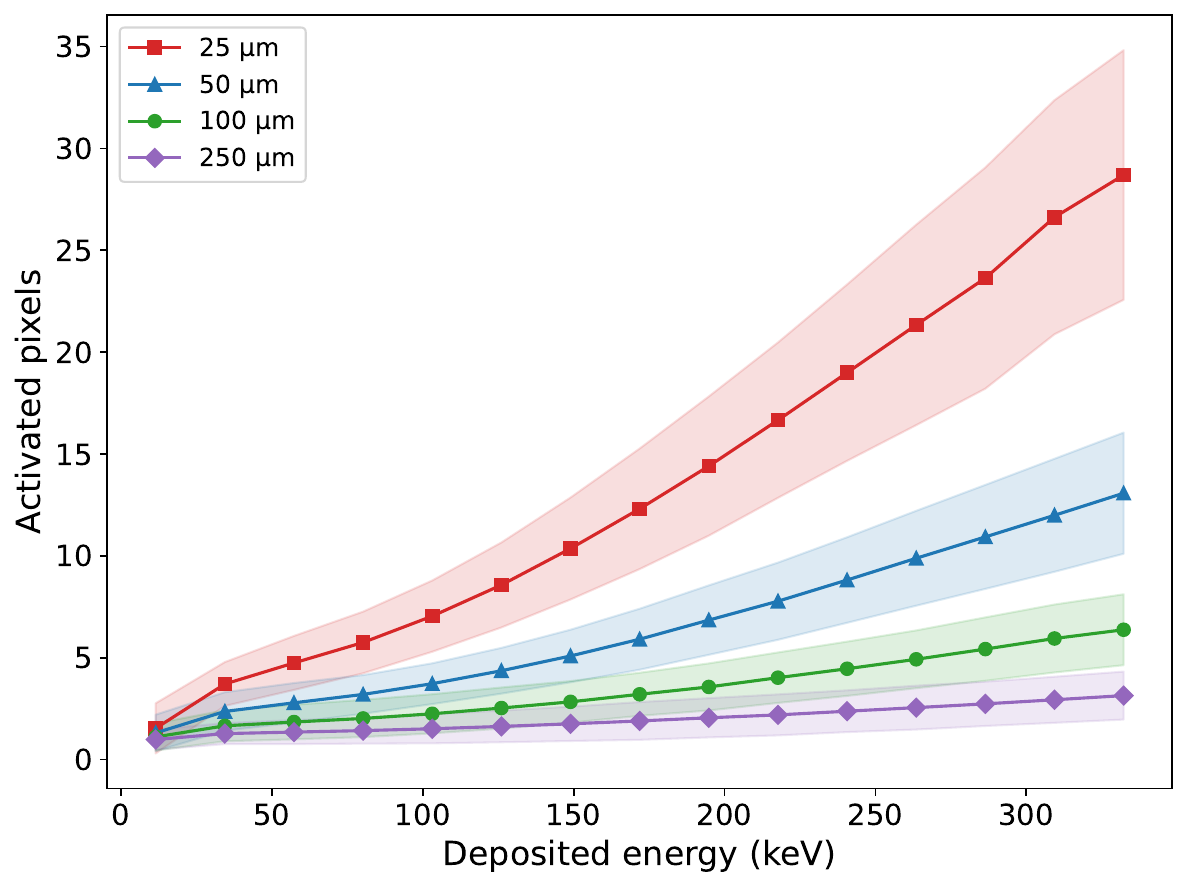}
\caption{Average number of activated pixels for different interaction deposited energies. Shaded area indicates the $1\sigma$ interval.}
\label{fig:pixels}
\end{figure}

Having established the optimal Gridsum~$N$ for each pixel size, we can now compare it to the Gaussian fit algorithm. The comparisons are shown in figure~\ref{fig:grid_vs_gauss}, and the average number of activated pixels is reported in figure~\ref{fig:pixels}.

The Gaussian fit worked well only if the energy deposition pattern in the detector followed a Gaussian distribution, which can be in first approximation true at low energies. The electron tracks became longer for higher energies, and the Gaussian character of the energy distribution was broadened or lost. The Gaussian fit accuracy sharply decreased at those energies, since the energy distribution no longer originated from charge sharing alone, but from a physical extension of the electron track. The Gridsum~$N$ algorithm accuracy remained more stable at higher energies where the Bragg peak was more defined, improving the spatial resolution compared to the Gaussian fit. The energy threshold above which Gridsum~$N$ outperformed the Gaussian fit was dependent on the pixel pitch: smaller pitches lowered this threshold, as the Bragg peak became spatially resolved across multiple pixels at lower interaction energies. For larger pixel pitches, the charge sharing was reduced and few pixels were activated, making the Bragg peak harder to isolate, raising the threshold. For the $250\,\mu$m pixel pitch case, the Gaussian fit outperformed Gridsum~$N$ for every $N$ tested in the energy range of interest.
 
\begin{figure}[htbp]
    \centering
    \begin{subfigure}{0.48\textwidth}
        \centering
        \includegraphics[width=\textwidth]{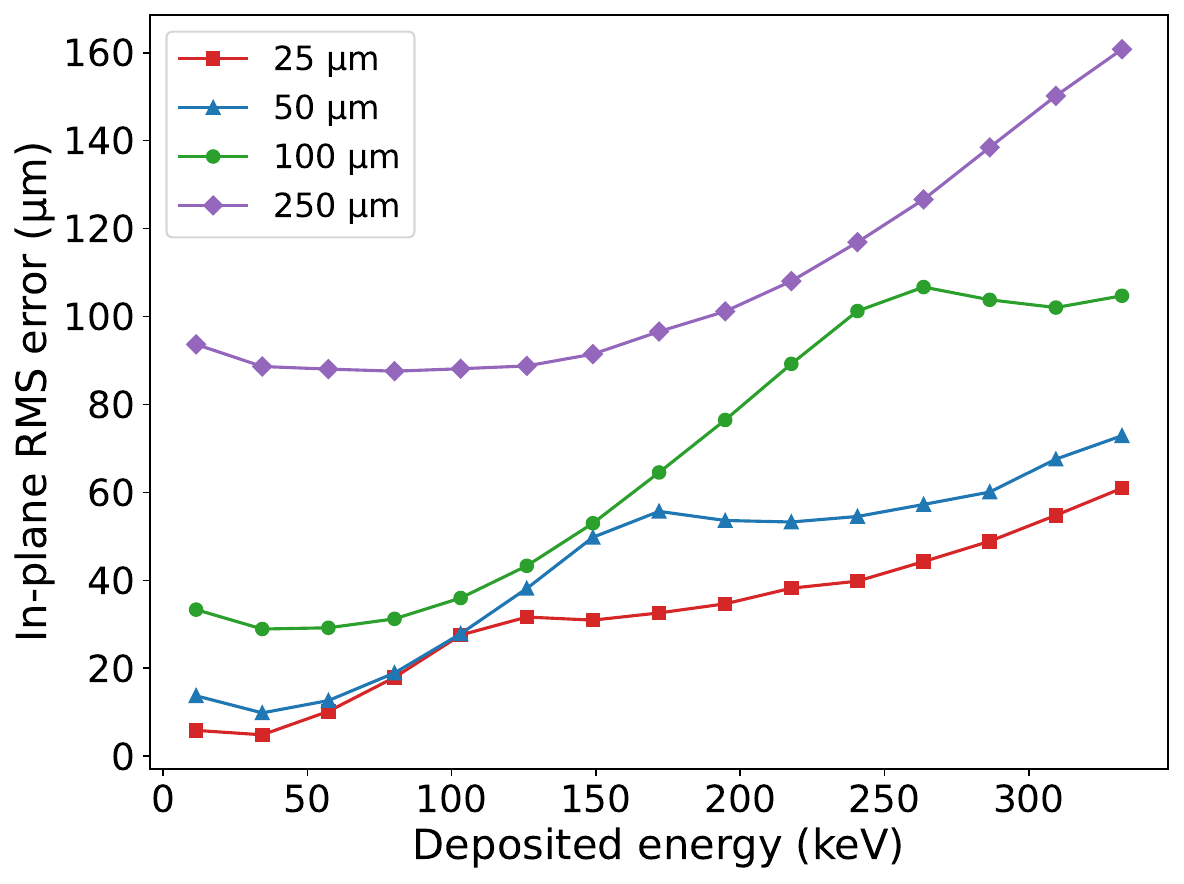}
        \caption{}
        \label{fig:xyres}
    \end{subfigure}
    \hfill
    \begin{subfigure}{0.48\textwidth}
        \centering
        \includegraphics[width=\textwidth]{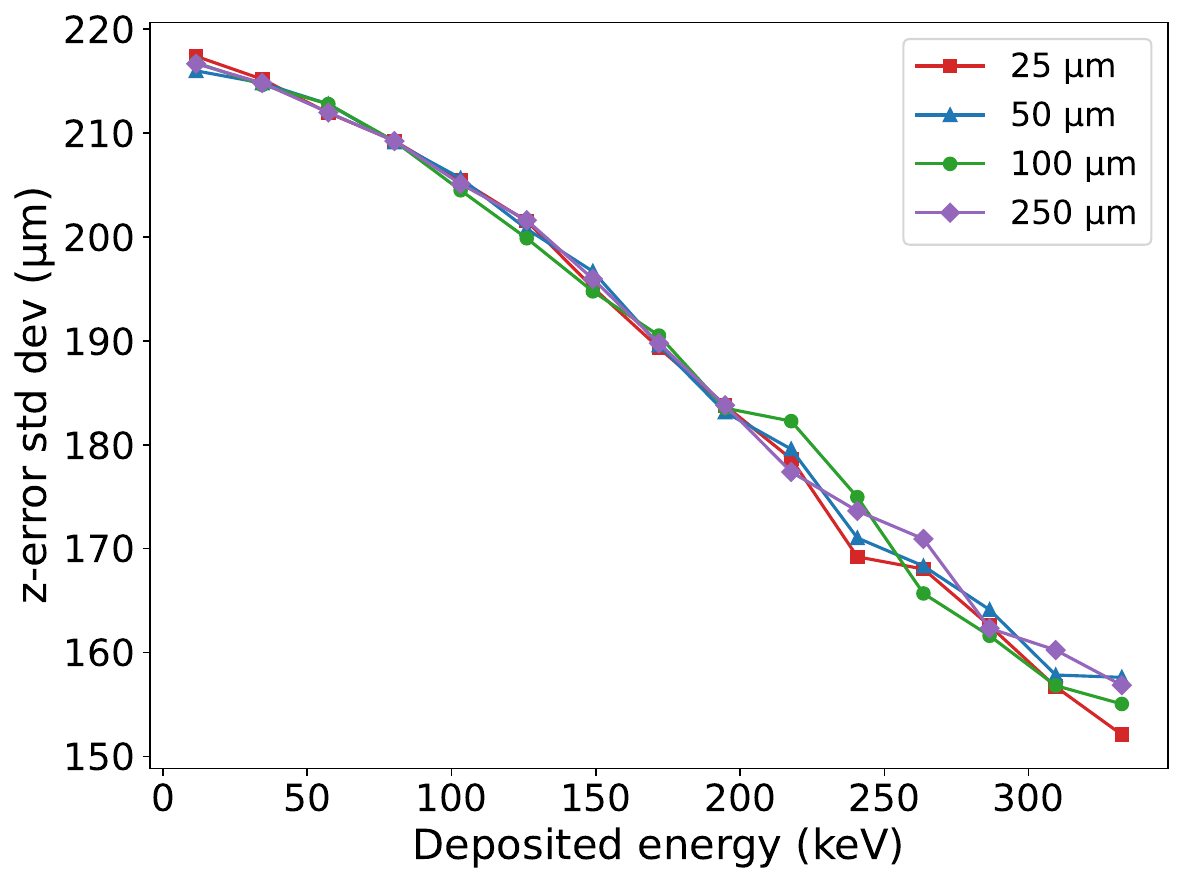}
        \caption{}
        \label{fig:zres}
    \end{subfigure}
    \caption{Minimized in-plane RMS error (a), and z direction error std (b) for different pixel pitches.}
    \label{fig:resolution_comparison}
\end{figure}

For each energy bin, the algorithm yielding the highest localization precision is selected, maximizing the overall in-plane spatial resolution across the full energy range and leading to the results shown in figure~\ref{fig:xyres}.

The spatial resolution results for the out-of-plane direction are reported in figure~\ref{fig:zres}. The algorithm used to estimate the $z$ interaction position is independent of pixel pitch, and this agrees well with the results shown in the plot. At low energies, the standard deviation fluctuates around 216\,$\mu$m, which agrees with the theoretical standard deviation of a uniform distribution over the 750\,$\mu$m sensor thickness ($\sigma = 750/\sqrt{12} \approx 216\,\mu$m), indicating that the $z$ position estimate carries little information beyond the sensor layer index. As the interaction energy increases, the recoil electron track extends over a greater range, making it increasingly likely to cross into an adjacent sensor layer, improving the depth estimate for a larger fraction of events and thus reducing the overall out-of-plane resolution. 
 
To estimate the overall performances for both energy sources and all pixel configurations simulated, we calculated the energy-averaged resolution as eq.~\eqref{eq:weighted_res}:

\begin{equation}
  R = \frac{\displaystyle\sum_{i} f_{i}\,R_{i}}{\displaystyle\sum_{i} f_{i}} \,,
  \label{eq:weighted_res}
\end{equation}

where $f_{i}$ are the counts in deposited energy bin $i$ from the source spectra in figure~\ref{fig:spectra} and $R_{i}$ is the resolution extracted from figure~\ref{fig:xyres} and figure~\ref{fig:zres} at the corresponding energy. The results are shown in table~\ref{tab:resolution}, where it can be noticed that the in-plane resolution scales approximately linearly with pixel pitch.

\begin{table}[htbp]
\centering
\caption{Summary of energy-averaged spatial resolution.}
\label{tab:resolution}
\begin{tabular}{cccc}
\hline
\textbf{Source energy} (keV) & \textbf{Pixel pitch} ($\mu$m) & \textbf{In-plane resolution} ($\mu$m) & \textbf{$z$ resolution} ($\mu$m) \\
\hline
 140 & 25  &  8.2 & 215 \\
 140 & 50  & 14.2 & 214 \\
 140 & 100 & 32.2 & 215 \\
 140 & 250 & 91.4 & 215 \\
 511 & 25  & 14.7 & 209 \\
 511 & 50  & 20.9 & 209 \\
 511 & 100 & 39.2 & 209 \\
 511 & 250 & 94.0 & 209 \\
\hline
\end{tabular}
\end{table}

% ---------------------------------------------------------------
\section{Conclusion}
\label{sec:conclusion}
% ---------------------------------------------------------------
 
In this work, we evaluated four different pixel pitches and two different algorithms to assess the interaction positions in a novel silicon detector. Monte Carlo simulations were performed for 140\,keV and 511\,keV photon sources embedded in a water phantom. The Gaussian fit algorithm based on charge sharing outperformed the Gridsum\,N algorithm at lower interaction energies. For higher interaction energies, regardless of pixel pitch, the Gridsum\,N approach exhibited better accuracy, thanks to more elongated tracks that allowed more easily to characterize the Bragg peak. The energy-weighted in-plane resolutions achieved, with exception for the 250\,$\mu$m case, were smaller than those reported for state-of-the-art Compton imaging detectors (see table~\ref{tab:comparison}), demonstrating the potential of this silicon geometry for high-resolution nuclear medicine applications. The out-of-plane resolution was found to be mostly limited by the sensor thickness, achieving energy weighted results around 209\,$\mu$m for 511\,keV photons and 215\,$\mu$m for 140\,keV ones. These results show that pixel pitches of around 100\,$\mu$m can achieve competitive interaction localization precision of tens of micrometers, while simultaneously providing sufficient electronics area to relax the constraints on the CMOS readout design. Future work will focus on optimizing the readout electronics design based on the trade-off between spatial resolution, power consumption, electronic integration area and complexity, validating the simulation results experimentally.

% ---------------------------------------------------------------
\section*{Code and data availability}
The configuration files used in this study to run Allpix$^2$ simulations are publicly available via Zenodo (https://doi.org/10.5281/zenodo.21623083). The COMSOL generated fields, the raw ROOT data, and the post-processing code are available upon request.

\acknowledgments
% ---------------------------------------------------------------
 
This research was supported by the Swedish Research Council (grant number 2022-03688\_VR) and the European Union (ERC, Si3, grant agreement 101142294). Views and opinions expressed are those of the authors only and do not necessarily reflect those of the European Union or the European Research Council Executive Agency. Neither the European Union nor the granting authority can be held responsible for them.
 
This work did not involve human subjects or animals in its research.
 
% ---------------------------------------------------------------

\bibliographystyle{unsrt}
\bibliography{references} 

@article{andreyev2010,
  title = {Reconstructed {{Image Spatial Resolution}} of {{Multiple Coincidences Compton Imager}}},
  author = {Andreyev, Andriy and Sitek, Arkadiusz and Celler, Anna},
  year = 2010,
  month = feb,
  journal = {IEEE Transactions on Nuclear Science},
  volume = {57},
  number = {1},
  pages = {151--159},
  issn = {1558-1578},
  doi = {10.1109/TNS.2009.2038053},
  urldate = {2026-01-29}
}

@article{caffrey2021,
  title = {Gamma-Ray Imaging Performance of the {{GRI}}+ {{Compton}} Camera},
  author = {Caffrey, A. and {Harkness-Brennan}, L.J. and Alshammari, H. and Boston, A.J. and Boston, H.C. and Griffiths, O. and Judson, D.S. and Kalantan, S. and Le Crom, B. and Nolan, P.J. and Powell, A. and Platt, J. and Randall, G. and Reid, C. and Rintoul, E. and Unsworth, C. and Woodroof, T.F. and Burrows, I. and Lazarus, I.H. and Simpson, J. and Atkinson, K.D.},
  year = 2021,
  month = sep,
  journal = {Journal of Instrumentation},
  volume = {16},
  number = {09},
  pages = {P09015},
  publisher = {IOP Publishing},
  issn = {1748-0221},
  doi = {10.1088/1748-0221/16/09/P09015},
  urldate = {2026-01-29},
  langid = {english}
}

@article{caravaca2022,
  title = {Compton and {{Proximity Imaging}} of {{Ac In Vivo With}} a {{CZT Gamma Camera}}: {{A Proof}} of {{Principle With Simulations}}},
  shorttitle = {Compton and {{Proximity Imaging}} of {{Ac In Vivo With}} a {{CZT Gamma Camera}}},
  author = {Caravaca, Javier and Huh, Yoonsuk and Gullberg, Grant T. and Seo, Youngho},
  year = 2022,
  month = nov,
  journal = {IEEE Transactions on Radiation and Plasma Medical Sciences},
  volume = {6},
  number = {8},
  pages = {904--915},
  issn = {2469-7303},
  doi = {10.1109/TRPMS.2022.3166116},
  urldate = {2026-01-29}
}

@article{danielsson2021,
  title = {Photon-Counting x-Ray Detectors for {{CT}}},
  author = {Danielsson, Mats and Persson, Mats and Sj{\"o}lin, Martin},
  year = 2021,
  month = jan,
  journal = {Physics in Medicine and Biology},
  volume = {66},
  number = {3},
  pages = {03TR01},
  issn = {1361-6560},
  doi = {10.1088/1361-6560/abc5a5},
  langid = {english},
  pmid = {33113525}
}

@article{darocharolo2025,
  title = {{{ARCADIA}} Fully Depleted {{CMOS MAPS}} Development with {{LFoundry}} 110 Nm {{CIS}}},
  author = {Da Rocha Rolo, Manuel and Andreazza, Attilio and Ambrosi, Giovanni and Alfonsi, Fabrizio and Balbi, Gabriele and Dalla Betta, Gian-Franco and Croci, Tommaso and Caccia, Massimo and Chiappara, Davide and Corradino, Thomas and Durando, Stefano and Falchieri, Davide and Gabrielli, Alessandro and Giubilato, Piero and Garbolino, Sara and Iuppa, Roberto and Mandurrino, Marco and Mattiazzo, Serena and Neub{\"u}ser, Coralie and Pancheri, Lucio and Pantano, Devis and Paterno, Andrea and Pezzoli, Matteo and Placidi, Pisana and Ratti, Lodovico and Ricci, Ester and Rivetti, Angelo and Santoro, Romualdo and Traversi, Gianluca and Wyss, Jeffery},
  year = 2025,
  month = aug,
  journal = {Frontiers in Sensors},
  volume = {6},
  publisher = {Frontiers},
  issn = {2673-5067},
  doi = {10.3389/fsens.2025.1603755},
  urldate = {2026-02-02},
  langid = {english}
}

@article{du2001,
  title = {Evaluation of a {{Compton}} Scattering Camera Using 3-{{D}} Position Sensitive {{CdZnTe}} Detectors},
  author = {Du, Y. F. and He, Z. and Knoll, G. F. and Wehe, D. K. and Li, W.},
  year = 2001,
  month = jan,
  journal = {Nuclear Instruments and Methods in Physics Research Section A: Accelerators, Spectrometers, Detectors and Associated Equipment},
  volume = {457},
  number = {1},
  pages = {203--211},
  issn = {0168-9002},
  doi = {10.1016/S0168-9002(00)00669-0},
  urldate = {2026-02-19}
}

@article{greffier2025,
  title = {Photon-Counting {{CT}} Systems: {{A}} Technical Review of Current Clinical Possibilities},
  shorttitle = {Photon-Counting {{CT}} Systems},
  author = {Greffier, Jo{\"e}l and Viry, Ana{\"i}s and Robert, Antoine and Khorsi, Mouad and {Si-Mohamed}, Salim},
  year = 2025,
  month = feb,
  journal = {Diagnostic and Interventional Imaging},
  volume = {106},
  number = {2},
  pages = {53--59},
  issn = {2211-5684},
  doi = {10.1016/j.diii.2024.09.002},
  urldate = {2026-02-23}
}

@article{griffiths2026,
  title = {Assessing the Impact of Position Resolution on Gamma-Ray Imaging Performance of a {{Segmented Inverted Coaxial Germanium}} ({{SIGMA}}) Detector},
  author = {Griffiths, Owain and Alharbi, Ahmed and Altasan, Nawaf and Das, Sneha and Harkness, Laura and Holloway, Fraser and Judson, Daniel and Richardson, Emily and Rintoul, Ellis},
  year = 2026,
  month = jan,
  journal = {Applied Radiation and Isotopes},
  volume = {227},
  pages = {112262},
  issn = {0969-8043},
  doi = {10.1016/j.apradiso.2025.112262},
  urldate = {2026-02-19}
}

@article{kim2024,
  title = {A Comprehensive Review on {{Compton}} Camera Image Reconstruction: From Principles to {{AI}} Innovations},
  shorttitle = {A Comprehensive Review on {{Compton}} Camera Image Reconstruction},
  author = {Kim, Soo Mee and Lee, Jae Sung},
  year = 2024,
  month = sep,
  journal = {Biomedical Engineering Letters},
  volume = {14},
  number = {6},
  pages = {1175--1193},
  issn = {2093-9868},
  doi = {10.1007/s13534-024-00418-8},
  urldate = {2026-02-23},
  pmcid = {PMC11502649},
  pmid = {39465108}
}

@article{krimmer2015,
  title = {Development of a {{Compton}} Camera for Medical Applications Based on Silicon Strip and Scintillation Detectors},
  author = {Krimmer, J. and Ley, J. -L. and Abellan, C. and Cachemiche, J. -P. and Caponetto, L. and Chen, X. and Dahoumane, M. and Dauvergne, D. and Freud, N. and Joly, B. and Lambert, D. and Lestand, L. and L{\'e}tang, J. M. and Magne, M. and Mathez, H. and Maxim, V. and Montarou, G. and Morel, C. and Pinto, M. and Ray, C. and Reithinger, V. and Testa, E. and Zoccarato, Y.},
  year = 2015,
  month = jul,
  journal = {Nuclear Instruments and Methods in Physics Research Section A: Accelerators, Spectrometers, Detectors and Associated Equipment},
  series = {New {{Developments}} in {{Photodetection NDIP14}}},
  volume = {787},
  pages = {98--101},
  issn = {0168-9002},
  doi = {10.1016/j.nima.2014.11.042},
  urldate = {2026-01-29}
}

@article{moser2009,
  title = {Silicon Detector Systems in High Energy Physics},
  author = {Moser, Hans-G{\"u}nther},
  year = 2009,
  month = jul,
  journal = {Progress in Particle and Nuclear Physics},
  volume = {63},
  number = {1},
  pages = {186--237},
  issn = {0146-6410},
  doi = {10.1016/j.ppnp.2008.12.002},
  urldate = {2026-02-23}
}

@inproceedings{olave2018,
  author        = {Olave, E.J. and Mattiazzo, S. and Panati, S. and Cossio, F. and Rivetti, A. and Pantano, D. and Demaria, N. and Pancheri, L. and Giubilato, P. and {Da Rocha Rolo}, M.D.},
  title         = {{MATISSE}: a low power front-end electronics for {MAPS} characterization},
  booktitle     = {Proceedings of Topical Workshop on Electronics for Particle Physics},
  series        = {PoS},
  volume        = {TWEPP-17},
  pages         = {016},
  year          = {2018},
  doi           = {10.22323/1.313.0016},
  publisher     = {SISSA Medialab},
}

@article{park2007,
  title = {Effects of {{Positron Range}} and {{Annihilation Photon Acolinearity}} on {{Image Resolution}} of a {{Compton PET}}},
  author = {Park, Sang-June and Rogers, W. Leslie and Clinthorne, Neal H.},
  year = 2007,
  month = oct,
  journal = {IEEE Transactions on Nuclear Science},
  volume = {54},
  number = {5},
  pages = {1543--1552},
  issn = {1558-1578},
  doi = {10.1109/TNS.2007.902358},
  urldate = {2026-01-29}
}

@article{plimley2011,
  title = {Reconstruction of Electron Trajectories in High-Resolution {{Si}} Devices for Advanced {{Compton}} Imaging},
  author = {Plimley, Brian and Chivers, Daniel and Coffer, Amy and Aucott, Tim and Wang, Wenni and Vetter, Kai},
  year = 2011,
  month = oct,
  journal = {Nuclear Instruments and Methods in Physics Research Section A: Accelerators, Spectrometers, Detectors and Associated Equipment},
  series = {Symposium on {{Radiation Measurements}} and {{Applications}} ({{SORMA}}) {{XII}} 2010},
  volume = {652},
  number = {1},
  pages = {595--598},
  issn = {0168-9002},
  doi = {10.1016/j.nima.2011.01.133},
  urldate = {2026-02-18}
}

@article{plimley2013,
  title = {Experimental {{Benchmark}} of {{Electron Trajectory Reconstruction Algorithm}} for {{Advanced Compton Imaging}}},
  author = {Plimley, Brian and Chivers, Daniel and Coffer, Amy and Vetter, Kai},
  year = 2013,
  month = jun,
  journal = {IEEE Transactions on Nuclear Science},
  volume = {60},
  number = {3},
  pages = {2308--2313},
  issn = {1558-1578},
  doi = {10.1109/TNS.2013.2254498},
  urldate = {2026-02-18}
}

@article{spannagel2018,
  title = {Allpix2: {{A}} Modular Simulation Framework for Silicon Detectors},
  shorttitle = {Allpix2},
  author = {Spannagel, S. and Wolters, K. and Hynds, D. and Alipour Tehrani, N. and Benoit, M. and Dannheim, D. and Gauvin, N. and N{\"u}rnberg, A. and Sch{\"u}tze, P. and Vicente, M.},
  year = 2018,
  month = sep,
  journal = {Nuclear Instruments and Methods in Physics Research Section A: Accelerators, Spectrometers, Detectors and Associated Equipment},
  volume = {901},
  pages = {164--172},
  issn = {0168-9002},
  doi = {10.1016/j.nima.2018.06.020},
  urldate = {2026-02-02}
}

@article{sundberg2021,
  title = {1-{$\mu$}m Spatial Resolution in Silicon Photon-Counting {{CT}} Detectors},
  author = {Sundberg, Christel and Persson, Mats U. and Wikner, J. Jacob and Danielsson, Mats},
  year = 2021,
  month = nov,
  journal = {Journal of Medical Imaging},
  volume = {8},
  number = {6},
  pages = {063501},
  publisher = {SPIE},
  issn = {2329-4302, 2329-4310},
  doi = {10.1117/1.JMI.8.6.063501},
  urldate = {2026-01-29}
}

@article{takeda2009,
  title = {Experimental {{Results}} of the {{Gamma-Ray Imaging Capability With}} a {{Si}}/{{CdTe Semiconductor Compton Camera}}},
  author = {Takeda, Shin'ichiro and Aono, Hiroyuki and Okuyama, Sho and Ishikawa, Shin-nosuke and Odaka, Hirokazu and Watanabe, Shin and Kokubun, Motohide and Takahashi, Tadayuki and Nakazawa, Kazuhiro and Tajima, Hiroyasu and Kawachi, Naoki},
  year = 2009,
  month = jun,
  journal = {IEEE Transactions on Nuclear Science},
  volume = {56},
  number = {3},
  pages = {783--790},
  issn = {1558-1578},
  doi = {10.1109/TNS.2008.2012059},
  urldate = {2026-01-29}
}

@article{turchetta2001,
  title = {A Monolithic Active Pixel Sensor for Charged Particle Tracking and Imaging Using Standard {{VLSI CMOS}} Technology},
  author = {Turchetta, R and Berst, J. D and Casadei, B and Claus, G and Colledani, C and Dulinski, W and Hu, Y and Husson, D and Le Normand, J. P and Riester, J. L and Deptuch, G and Goerlach, U and Higueret, S and Winter, M},
  year = 2001,
  month = feb,
  journal = {Nuclear Instruments and Methods in Physics Research Section A: Accelerators, Spectrometers, Detectors and Associated Equipment},
  volume = {458},
  number = {3},
  pages = {677--689},
  issn = {0168-9002},
  doi = {10.1016/S0168-9002(00)00893-7},
  urldate = {2026-02-23}
}

@article{turecek2018,
  title = {Compton Camera Based on {{Timepix3}} Technology},
  author = {Turecek, D. and Jakubek, J. and Trojanova, E. and Sefc, L.},
  year = 2018,
  month = nov,
  journal = {Journal of Instrumentation},
  volume = {13},
  number = {11},
  pages = {C11022},
  issn = {1748-0221},
  doi = {10.1088/1748-0221/13/11/C11022},
  urldate = {2026-04-22},
  langid = {english}
}

@article{vanaudenhaege2015,
  title = {Review of {{SPECT}} Collimator Selection, Optimization, and Fabrication for Clinical and Preclinical Imaging},
  author = {Van Audenhaege, Karen and Van Holen, Roel and Vandenberghe, Stefaan and Vanhove, Christian and Metzler, Scott D. and Moore, Stephen C.},
  year = 2015,
  month = aug,
  journal = {Medical Physics},
  volume = {42},
  number = {8},
  pages = {4796--4813},
  issn = {0094-2405},
  doi = {10.1118/1.4927061},
  urldate = {2026-02-23},
  pmcid = {PMC5148182},
  pmid = {26233207}
}

@article{vetter2007,
  title = {High-Sensitivity {{Compton}} Imaging with Position-Sensitive {{Si}} and {{Ge}} Detectors},
  author = {Vetter, K. and Burks, M. and Cork, C. and Cunningham, M. and Chivers, D. and Hull, E. and Krings, T. and Manini, H. and Mihailescu, L. and Nelson, K. and Protic, D. and Valentine, J. and Wright, D.},
  year = 2007,
  month = aug,
  journal = {Nuclear Instruments and Methods in Physics Research Section A: Accelerators, Spectrometers, Detectors and Associated Equipment},
  series = {Proceedings of the 11th {{Symposium}} on {{Radiation Measurements}} and {{Applications}}},
  volume = {579},
  number = {1},
  pages = {363--366},
  issn = {0168-9002},
  doi = {10.1016/j.nima.2007.04.076},
  urldate = {2026-01-29}
}

@article{yabu2022,
  title = {Tomographic {{Imaging}} by a {{Si}}/{{CdTe Compton Camera}} for {$^{111}$}{{In}} and {$^{131}$}{{I Radionuclides}}},
  author = {Yabu, Goro and Yoneda, Hiroki and Orita, Tadashi and Takeda, Shin'ichiro and Caradonna, Pietro and Takahashi, Tadayuki and Watanabe, Shin and Moriyama, Fumiki},
  year = 2022,
  month = may,
  journal = {IEEE Transactions on Radiation and Plasma Medical Sciences},
  volume = {6},
  number = {5},
  pages = {592--600},
  issn = {2469-7303},
  doi = {10.1109/TRPMS.2021.3104665},
  urldate = {2026-01-29}
}

@article{yoon2020,
  title = {Estimate of the {{225Ac Radioactive Isotope Distribution}} by {{Means}} of {{DOI Compton Imaging}} in {{Targeted Alpha Radiotherapy}}: {{A Monte Carlo Simulation}}},
  shorttitle = {Estimate of the {{225Ac Radioactive Isotope Distribution}} by {{Means}} of {{DOI Compton Imaging}} in {{Targeted Alpha Radiotherapy}}},
  author = {Yoon, Changyeon and Jo, Seongmin and Cho, Yongho and Kim, Nakjeom and Lee, Taewoong},
  year = 2020,
  month = may,
  journal = {Journal of the Korean Physical Society},
  volume = {76},
  number = {10},
  pages = {954--960},
  issn = {1976-8524},
  doi = {10.3938/jkps.76.954},
  urldate = {2026-01-29},
  langid = {english}
}
 
\end{document}